\documentclass[twocolumn,showpacs,keywords,preprintnumbers, prc,superscriptaddress]{revtex4}
\usepackage{epsfig}
\usepackage{graphicx}
\usepackage{epstopdf}
\usepackage{amsmath,amssymb,amsfonts}
\usepackage{array}
\usepackage{url}
\usepackage{hyperref}
\usepackage{lineno}
\usepackage{xspace}
\usepackage[usenames,dvipsnames]{color}
\usepackage[toc,page]{appendix}
\usepackage{float}

\newcommand{\sNN}{{{$\sqrt{s_{_{\mathrm{NN}}}}$}}}

\newcommand{ \be }{\begin{equation}}
\newcommand{ \ee }{\end{equation}}

\newcommand{\bef}{\begin{figure}}
\newcommand{\eef}{\end{figure}}
\newcommand{\bc}{\begin{center}}
\newcommand{\ec}{\end{center}}

\newcommand{\auau}{\mbox{Au$+$Au}\xspace}

\begin{document}
\title{Effect of limited statistics on higher order cumulants measurement\\ in heavy-ion collision experiments.}

\author{Ashish Pandav}
\email{ashish.pandav@niser.ac.in} 

\author{Debasish Mallick}
\email{debasish.mallick@niser.ac.in}  

\author{Bedangadas Mohanty} 
\email{bedanga@niser.ac.in} 

\affiliation{School of Physical Sciences, National Institute of Science Education and Research, HBNI, Jatni 752050, India}

\begin{abstract}
We have studied the effect of limited statistics of data on measurement of the different order of cumulants of net-proton distribution assuming that the proton and antiproton distributions follow Possionian and Binomial distributions with initial parameters determined from experimental results for two top center of mass energies (\sNN$=200$ and $62.4$~GeV) in most central ($0-5$\%) Au$+$Au collisions at Relativistic Heavy Ion Collider (RHIC). In this simulation, we observe that the central values for higher order cumulants have a strong dependence on event sample size and due to statistical randomness the central values of higher order cumulants could become negative. We also present a study on the determination of the statistical error on cumulants using delta theorem, bootstrap and sub-group methods and verified their suitability by employing a Monte Carlo procedure. Based on our study we find that the bootstrap method provides a robust way for statistical error estimation on high order cumulants. We also present the exclusion limits on the minimum event statistics needed for determination of cumulants if the signal strength (phase transition or critical point) is at a level of $5$\% and $10$\% above the statistical level. This study will help the experiments to arrive at the minimum required event statistics and choice of proper method for statistical error estimation for high order cumulant measurements.
\end{abstract}
\maketitle

\section{Introduction}
Relativistic collision of heavy nuclei, by varying center of mass collision energy (\sNN), allows us to map the quantum chromodynamics (QCD) phase diagram characterized by temperature $T$ and baryonic chemical potential $\mu_{B}$ ~\cite{Mohanty:2009vb, SN0493,SN0598, starNP10, starNP,starNC, starNK,BraunMunzinger:2007zz}. Lattice QCD calculations suggest a smooth crossover transition from hadronic matter (a state of confined quarks and gluons) to a deconfined state of quarks and gluons for $\mu_{B}=0$ at certain finite $T$~\cite{Aoki:2006we}. Also at larger value of $\mu_{B}$, theoretical QCD model predictions of a first-order phase boundary~\cite{Ejiri:2008xt, Bowman:2008kc} and existence of the QCD critical point~\cite{qcp, qcp1} are the motivations for carrying out high energy heavy-ion collision with varying beam energies. Higher order cumulants of conserved quantities such as the net-baryon, net-charge, and net-strangeness numbers, as a function of beam energy, are expected to show non-monotonic behaviour near the critical point~\cite{Gupta:2011wh,stephanovmom}. Near the QCD critical point, the third order cumulant (or skewness) and fourth order cumulant (or kurtosis) of net-baryon are expected to be negative valued~\cite{Asakawa:2009aj, Stephanov:2011pb} while the sixth order cumulant of both net-baryon and net-charge distribution is expected to be negative near a crossover phase transition~\cite{negc6:friman}. Using heavy-ion collider facility at RHIC, STAR experiment has performed measurements on higher order moments of net-proton~\cite{starNP10, starNP}, net-charge~\cite{starNC} and net-kaon~\cite{starNK} multiplicity distributions. PHENIX experiment at RHIC also has carried out measurements of moments of net-charge distributions~\cite{phenixNC}. At the Large Hadron Collider (LHC), ALICE experiment is also studying higher order moments to characterize the nature and order of the QCD phase transition~\cite{alice:fluct1, alice:meanpt, alice:np}.

Higher order cumulants of net-baryon and net-charge fluctuations are predicted to be sensitive to the nature of the QCD transition (crossover and chiral). In particular, ratios of the sixth to second and eighth to second order cumulants of the net-baryon number fluctuations change rapidly and the sixth order cumulants of both net-baryon and net-charge are predicted to remain negative if the system formed in such collisions freeze-out in the proximity of crossover region of the QCD phase diagram~\cite{negc6:friman, negc6:cheng}. The STAR experiment reported a preliminary result on centrality dependence of $C_{6}/C_{2}$ of net-proton and net-charge distribution in \auau collisions at \sNN$=200$~GeV~\cite{Esha:2017yeh}. $C_{6}/C_{2}$ of net-proton distribution exhibits $-ve$ values systematically across peripheral ($70-80\%$) to most central ($0-5\%$) collisions and net-charge results shows $-ve$ values only for most central ($0-5\%$) collisions. Further STAR experiment has reported preliminary results that reveal a non-monotonic variation of ratio of fourth order cumulant to second order cumulant of net-proton distributions ($C_{4}/C_{2}$) with beam energy~\cite{luo:NP}. This trend in measurement is similar to as expected, if the system traverses in the vicinity of a critical point in the QCD phase diagram. However due to finite size and time effects the signals corresponding to phase transition and critical point as discussed above are expected to be small. A high statistics run as part of the second phase of the Beam Energy Scan program at RHIC is envisaged to make a statistically more accurate measurement of $C_{4}/C_{2}$~\cite{SN0493, SN0598}. \\

In this work, via Monte Carlo simulation, we investigated the effect of limited statistics on the values of higher order cumulants up to $C_{7}$ within two statistical distribution models. In the first model, net-proton distribution is taken as a Skellam distribution (assuming proton and antiproton distributions are Poissonian) where the input parameters are mean ($C_{1}$) of experimentally measured proton and antiproton distributions from STAR experiment. In the second model, net-proton is obtained from the difference of binomially distributed protons and antiprotons, where the experimental value of mean ($C_{1}$) and variance ($C_{2}$) of proton and antiproton are used to characterize the input binomial distributions. We have performed this study for two top energies (\sNN$=200$ and $62.4$~GeV) of heavy-ion (Au) collision at RHIC, the energy region where signals of crossover transition are predicted to manifest. However the conclusions of the work remain valid for both higher and lower beam energies. In this study, we see that the estimated higher order cumulants, especially in the case of relatively small event sample size, randomly fluctuate around their true values, saturate and approach their true values with larger event sample size (quantitative description of smaller and larger sample size varies with different aspects of simulation such as input models, input parameters etc. and depends on order of cumulants, hence the discussions are left to the later sections). The value of $C_{6}$ of net-proton as a function of sample size in certain cases is observed to be $-ve$ in both the collision energies due to limited statistics of sample used and approaches the $+ve$ true value as the sample size is increased. The statistical errors which should reflect the consistency of estimated results with their true values, are also studied. We investigate the robustness of commonly followed methods for statistical error estimation such as delta theorem~\cite{adasgupta:book}, bootstrap method~\cite{bootstrap, bootstrap1}, and sub-group method (used in some experiments)~\cite{Abdelwahab:2014yha, Acharya:2017cpf}. Finally we obtain the exclusion curves for minimum statistics required to estimate the cumulants of net-proton distribution with a precision of $5\%$ and $10\%$ (relative to their true values) for various scenarios within the ambit of the models studied here.

We organize the content of this work as follows. In the following section, we describe mathematical prescription of relationships between moments and cumulants of various orders, along with discussions on models, input parameters, and cumulants calculation. A brief discussion on error estimation methods is also included in this section. In section-III, the results on sample size dependence of various order net-proton cumulants and studies on statistical error estimation using various methods are presented. Interpretation and discussion on the significance of these results in the context of precision measurement of higher order cumulants for the experimental search of phase transition and critical point signals are carried out in the Section-IV. Finally, in Section-V we summarize the findings of this study.

\section{Methodology}
We have performed Monte Carlo simulations to study the effect of limited statistics on higher order cumulants. In the simulation, two statistical models are used to generate net-proton distributions with experimental inputs from two collision energies. In the first case, net-proton distribution is generated from the difference of binomially distributed protons and antiprotons characterized by mean ($1^{st}$ order cumulant) and variance ($2^{nd}$ order cumulant) of proton and antiproton measured in STAR experiment~\cite{starNP}. The proton and antiproton distributions are independently (no correlations) generated. In the second case, net-proton follows a Skellam distribution and experimental values of mean of proton and antiproton are used as input parameters for the simulation.
\subsection{Cumulants Calculation}
The higher order fluctuations are often expressed in terms of cumulants or moments.
Cumulants and central moments are related to each other in the following way.
Let the deviation of $N$ (any entry in the data sample) from its mean value ($\mu=<N>$, referred as $1^{st}$ raw moment) be defined by
\begin{equation}
\delta N=N-<N>.
\end{equation}
Any $r^{th}$ order central moment is defined as:
\begin{equation}
\mu_r = \,< (\delta N)^r >, 
\end{equation}
$i.e.$ $1^{st}$ order central moment turns out to be zero. $\mu _1=\,< (\delta N) >=0$.
The Cumulants of a given data sample could be written in terms of moments as follows.
\begin{eqnarray}
 C_1 &=&\mu \\
 C_2 &=&\mu_2  \\
 C_3 &=& \mu_3  \\
 C_n(n>3)  &=& \mu_n  - \sum\limits_{m = 2}^{n - 2}
{\left(
\begin{array}{l}
 n - 1 \\
 m - 1 \\
 \end{array} \right)C_m } \mu_{n - m}
\end{eqnarray}
For two independent variables $X$ and $Y$, the cumulants of probability distribution of their sum ($X+Y$), are just addition of cumulants of individual distributions for $X$ and $Y$, $i.e.$ $C_{n,X+Y}=C_{n,X}+C_{n,Y}$ for $n^{th}$ order cumulant. For distribution of difference in $X$ and $Y$, even order cumulants are addition of the individual cumulants, while odd order cumulants are obtained by subtracting the individual cumulants. Explicit expressions for cumulants in terms of central moments are mentioned in Appendix~\ref{sec:appA}.

\subsubsection*{Binomial and Poissonian Model}
In heavy-ion collisions, several physics processes could lead to the correlated production of protons and antiprotons. For simplicity, we do not include any physics motivated correlations between proton and antiproton multiplicity distributions. In the first case, we take protons and antiprotons generated independently each following a Poisson distribution. The probability distribution of a Poisson distributed discrete random variable $X$ is given as follows.
\begin{equation}
p(k;\lambda)=\mathrm{Pr}(X=k)=\frac{\lambda^{k}e^{-\lambda}}{k!},
\end{equation}
where $\lambda$ is the mean, $e$ is the Euler's number. The resultant distribution of difference in two Poisson distributed variables is a Skellam distribution.
The probability mass function for the Skellam distribution is given by,
\begin{equation}
p(k;\mu_{1},\mu_{2})=e^{-(\mu_{1}+\mu_{2})}\,\Big(\,\frac{\mu_{1}}{\mu_{2}}\,\Big)^{\,k/2}\,\,I_{k}  (2\sqrt{\mu_{1}\mu_{2}}),
\end{equation}
where $\mu_{1}$ and $\mu_{2}$ are the mean of Poisson distributions and $I_{k}(z)$ is the modified Bessel function of the first kind. This representation with above-specified inputs is referred to \emph{Poissonian} model in the text that follows. For simulation, experimental values of $C_{1}$ for proton and antiproton distributions are taken as the input parameters $\mu_{1}$ and $\mu_{2}$  respectively. \\ \\

Similarly, in the \emph{Binomial} model, we consider protons and antiprotons are independently generated each following a binomial distribution. The probability mass function of a binomially distributed random variable $k$ is given as follows.
\begin{equation}
B(k;n,p)=\frac{n!}{k!\,(n-k)!}\,p^{k}\,(1-p)^{k},
\end{equation}
where $n$ is the number of trials, $p$ is probability of success and $k$ is the number of sucesses.
Two parameters that characterize the binomial distribution are determined from relations such as $\mu=np$, $\sigma^{2}=np(1-p)$ where $\mu$ and $\sigma^{2}$ are the $1^{st}$ order ($C_{1}$) and $2^{nd}$ order cumulant ($C_{2}$) of proton or antiproton distributions measured in the experiment. Mathematical expressions for cumulants of Binomial and Skellam distributions are mentioned in Appendix~\ref{sec:appB}.
\\ \\
In this study experimental data used as inputs for simulation are (anti) proton cumulants in a kinematic region of $|y|<0.5$ and $0.4<p_{T}\mathrm{(GeV/c)}<0.8$ for most central (0-5\%) collisions measured by STAR collaboration~\cite{starNP}. Input parameters set from two energies $\sqrt{s_{\mathrm{NN}}}(\mathrm{GeV})=62.4, 200$ are considered for both the models. In all the cases studied, the maximum statistics is $10^{10}$ events. We study the effect of the size of an event sample on net-proton cumulants up to $7^{th}$ order for both the Binomial and Poisson models. The results and detailed discussion occur in the later sections.
 
\subsection{Error Estimation}
The higher order cumulants are sensitive to the shape of the distribution specially the tails, hence estimating the statistical uncertainties correctly is crucial. Further, in higher order cumulants analysis, the concern in the context of statistical error estimation has gone beyond the basic necessity of constraining an experimentally observed parameter in certain numerical range, to the affair of finding and testing of an accurate and reliable method for error estimation itself. We discuss here three such methods to estimate the statistical errors on cumulants: Delta theorem method~\cite{adasgupta:book, Luo:2011tp}, Bootstrap method~\cite{bootstrap, bootstrap1} and Sub-group method ~\cite{Abdelwahab:2014yha, Acharya:2017cpf}. Errors obtained are further subjected to a Monte Carlo verification procedure to test the reliability of the methods.
\subsubsection*{Delta theorem method}
Using delta theorem, a concise form of standard error propagation method, analytical formulae for statistical errors on cumulants and moments exist in the literature~\cite{Luo:2011tp, Luo:2014rea}. Delta theorem method has been quite extensively used in estimating statistical errors in the analysis of higher order cumulants in heavy-ion collision experiments~\cite{starNP10, starNP, starNC, starNK, phenixNC}. The error estimation method uses the Central Limit Theorem (CLT) - an important theorem on convergence of probability. One of the most common statements of the CLT is as follows.\\

\emph{Central Limit Theorem}: Suppose ${X_{1},X_{2}, .., X_{n}}$ is a collection of random variables that are independent and identically distributed with mean E$[X_{i}]=\mu$ and Var$[X_{i}]=\sigma^{2}<\infty$, then as $n$ approaches infinity, the random variable $\sqrt{n}(S_{n}-\mu)$ converges to a normal distribution $N(0,\sigma^{2})$, where $S_{n}=(X_{1}+X_{2}+..+X_{n})/n$. In other words, for large value of $n$, $S_{n}$ approximately follows a normal distribution $N(\mu,\sigma^{2}/n)$.\\ \\

Given the knowledge of the approximate distribution of a statistic itself, delta theorem gives a prescription to approximate the distribution of a transformation of the statistic in large samples~\cite{adasgupta:book}.
The statement of delta theorem is mentioned below.\\ \\
\textbf{Delta theorem :} Let $T_{n}$ be a sequence of statistics such that
\begin{equation}
\sqrt{n}(T_{n}-\theta)\,\, \lim_{\rightarrow} \,\, N(0,\sigma^{2}(\theta)), \, \, \sigma(\theta)>0. \\ 
\end{equation}
Let $g$ be a real function which atleast is differentiable at $\theta$ with $g^{'}(\theta)\neq 0$. Then \\
\begin{equation}
\sqrt{n}(g(T_{n})-g(\theta))\,\, \lim_{\rightarrow} \,\, N(0,[g^{'}(\theta)]^{2}\sigma^{2}(\theta)).
\end{equation}
Error on cumulants upto $8^{th}$ order expressed in terms of central moments are mentioned in Sec.~\ref{sec:appC} of Appendix~\ref{sec:appC}. 
Here, it is important to note that, the estimation of error on moments calculated from a sample requires the knowledge of corresponding parameters ($e.g.$ $\sigma$ in the statement of CLT and $\theta$ in the definition of delta theorem) of the population, which is never within the scope of comprehension in the experiments. Hence the moments calculated from the sample, are used as estimators for corresponding parameters of the population.
\subsubsection*{Bootstrap Method}
The bootstrap method finds the error on the estimators in an efficient Monte Carlo way by forming bootstrap samples without involving the complexities of standard error propagation method. It makes use of random selection of elements with replacement from the original sample, to construct bootstrap samples over which the sampling variance of an estimator is calculated~\cite{bootstrap, bootstrap1}.\\ \\
To start with, let $X$ be a random sample representing the experimental dataset drawn randomly from an unknown parent distribution. Let $\hat{e}$ be the estimator of a statistic (such as mean or variance), on which we intend to find the standard error. The sequence of steps followed to estimate standard error using the bootstrap method is as follows. \\ \\
1. Given a parent sample of size $n$, construct $B$ number of independent bootstrap samples $X^{*}_{1}$, $X^{*}_{2}$, $X^{*}_{3}$, ..., $X^{*}_{B}$, each consisting of $n$ data points randomly drawn with replacement from the parent sample. \\
2. Evaluate the estimator in each bootstrap sample,
\begin{equation}
\hat{e}^{*}_{b}=\hat{e}(X^{*}_{b})      \qquad b=1,2,3, ..., B.
\end{equation}
3. The sampling variance of the estimator is given as follows.
\begin{equation}
Var(\hat{e}) = \frac{1}{B-1}\sum_{b=1}^{B}\Big(\hat{e}^{*}_{b} - \bar{\hat{e}}\Big)^{2},
\end{equation}
where $\bar{\hat{e}}^{*} = \frac{1}{B}\sum_{b=1}^{B}(\hat{e}^{*}_{b})$\\ 
The sufficient enough value of $B$ for an accurate estimation of error within the bootstrap method varies from case to case depending upon the initial sample size. However, in general, the larger value of $B$ estimates the error better. We find the error on cumulants using the bootstrap method and the method is subjected to verification procedure to test the robustness. To determine the value of $B$ and maximize stability of estimated error with respect to the number of bootstrap samples, we further employ a qualitative test as discussed in results section~\ref{sec:Results}.
\subsubsection*{Sub-group Method}
Random replication of samples could also be done by dividing the initial data sample into many groups. To be more precise, without replacement of entries, subgroup method divides the data sample into sub-samples, evaluates the estimator over each sub-sample and estimates the sampling variance on the estimator. We mention the steps involved in the sub-group method as follows. \\ \\
Given the data sample $X$ and estimator of interest be denoted by $\hat{e}$, 

1. Construct $S$ number of sub-samples $X^{*}_{1}$, $X^{*}_{2}$, $X^{*}_{3}$, ..., $X^{*}_{S}$, each consisting of $n = \frac{N}{S}$ number of data points randomly drawn from the parent sample without replacement, where $N$ is the size of parent sample. \\

2. Calculate the value of estimator for each sub-sample,
\begin{equation}
\hat{e}^{*}_{s} = \hat{e}(X^{*}_{s})      \qquad s = 1,2,3, ..., S.
\end{equation}

3. Variance on the estimator is calculated as follows.
\begin{equation}
Var(\hat{e}) = \frac{1}{S(S-1)}\sum_{s=1}^{S}\Big(\hat{e}^{*}_{s} - \bar{\hat{e}}\Big)^{2},
\label{eq:subgrp_var}
\end{equation}
where $\bar{\hat{e}}^{*} = \frac{1}{S}\sum_{s=1}^{S}(\hat{e}^{*}_{s})$. The extra factor of $1/S$ in Eq.\ref{eq:subgrp_var} is used to scale back the variance on estimator from sub-sample level to original data sample level to account for the statistics mismatch.\\
We obtain the error on cumulants using the sub-group method for different sub-sample sizes and compared to the corresponding results from bootstrap and delta-theorem method. Gaussian behaviour of estimators from the sub-samples in sub-group method is tested by subjecting to a verification procedure.
\section{Results\label{sec:Results}}
\begin{figure*}[hbtp]
\centering 
\includegraphics[scale=0.3]{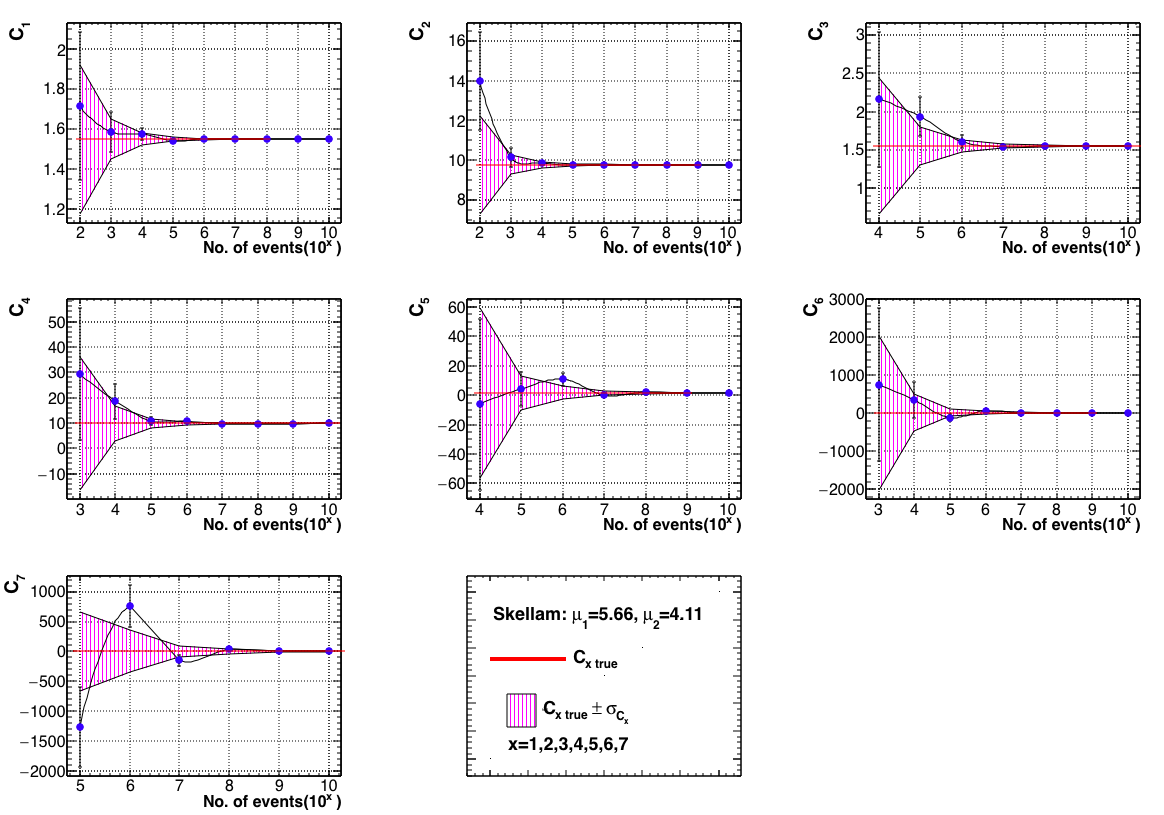}
\caption{(Color online) Cumulants of net-proton distribution upto $6^{th}$ order are shown as function of sample size. The net-proton distribution is a Skellam distribution with input parameters $\mu_{1}$ (and $\mu_{2}$) taken from experimental value of mean ($C_{1}$) of proton (and antiproton) distributions in most central (0-5\%) Au$+$Au collision at \sNN$=200$~GeV. Blue solid circle markers represent the calculated cumulant values from simulated net-proton distribution. The solid red line represents the true value of cumulants obtained using analytical formula for a Skellam distribution. The magenta band represents $\pm1\sigma$ statistical error (from delta theorem method) about the true value of cumulants.}
\label{fig:band_sklm_200G}
\end{figure*}
To find the effect of limited statistics in the measurement of higher order cumulants, we calculate cumulants of various order of simulated net-proton distributions within two models namely, Poissonian model and Binomial model. Statistical errors are estimated using delta theorem method, bootstrap method and sub-group method. We present the results on calculating back the cumulants values viz-a-viz the true values as a function of event statistics in the first part of this section and results on error estimation methods are investigated in the later part.
\subsubsection{\label{CumulantResults}Dependence of cumulants on sample size}
\begin{figure*}[hbtp]
\centering 
\includegraphics[scale=0.3]{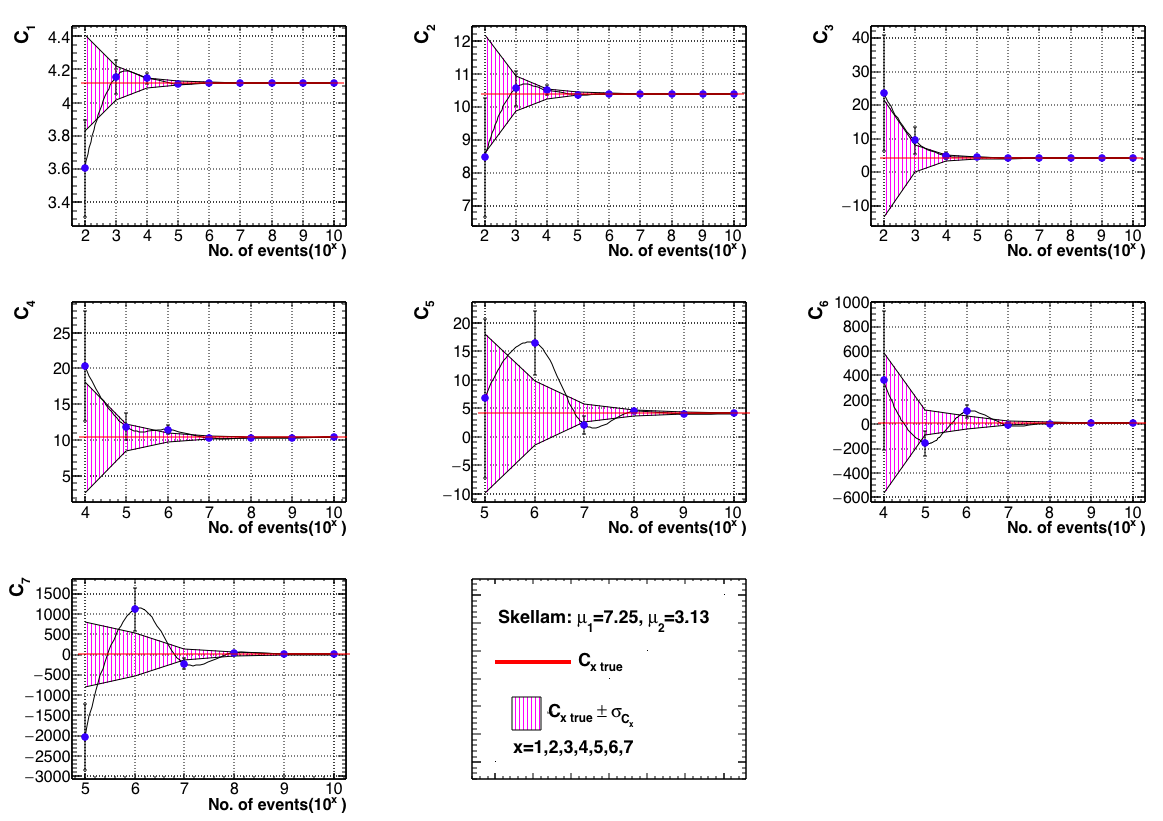}
\caption{(Color online) Cumulants of net-proton distribution upto $6^{th}$ order are shown as function of sample size. The net-proton distribution is a Skellam distribution with input parameters $\mu_{1}$ (and $\mu_{2}$) taken from experimental value of means ($C_{1}$) of proton (and antiproton) distributions in most central (0-5\%) Au$+$Au collision at \sNN$=200$~GeV. Blue solid circle markers represent the calculated cumulant values from simulated net-proton distribution. The solid red line represents the true value of cumulants obtained using analytical formula for a Skellam distribution. The magenta band represents $\pm1\sigma$ statistical error (from delta theorem method) about the true value of cumulants.}
\label{fig:band_sklm_62G} 
\end{figure*}
\begin{figure}[hbtp]
\includegraphics[scale=0.45]{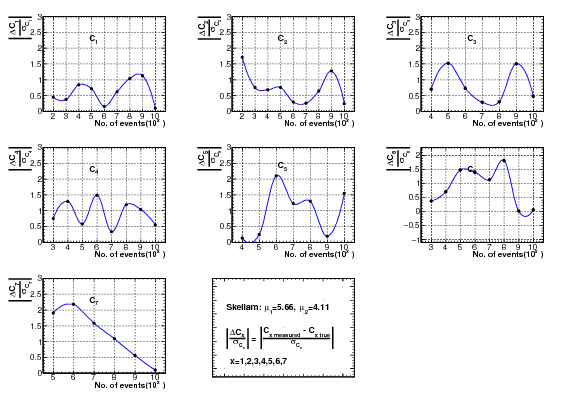}
\caption{(Color online) Sample size dependence of standard deviation (Modulus deviation of cumulants from their true values scaled with $1\sigma$ statistical error) of cumulants up to $7^{th}$ order in the Poissonian model with input parameters from proton and antiproton distribution in most central (0-5\%) Au$+$Au collision at \sNN$=200$~GeV.}
\label{fig:skell_chisq_200G} 
\end{figure}

Net-proton following a Skellam distribution is generated with input parameters $\mu_{1}=5.66$ and $\mu_{2}=4.11$ which are efficiency corrected mean of proton and antiproton distributions respectively, in most central (0-5\%) Au$+$Au collision at \sNN$=200$~GeV in STAR experiment~\cite{starNP}. As shown in Fig.~\ref{fig:band_sklm_200G}, values of cumulants (upto $7^{th}$ order) randomly fluctuate for the smaller value of sample statistics and quality of agreement with true values improves by increasing number of events in a sample. However, in most of the cases, as shown in Fig.~\ref{fig:skell_chisq_200G} the cumulants agree with their true values within $\pm 1\sigma$ statistical error while in very few cases cumulants fluctuate beyond $\pm 1\sigma$ error. To be precise, value of cumulants deviate in a range of $0.005\sigma$ to $2.1\sigma$. The fact that cumulants in some cases lie beyond $\pm 1\sigma$ error could be accounted to the small yet significant probability outside $\pm 1\sigma$ region of a Gaussian distribution. In few cases, $C_{5}$, $C_{6}$ and $C_{7}$ show negative values. Note that, sign of cumulants have physical significance in the field of studying QCD phase diagram. The statistical error is estimated using the delta theorem method. The true values of cumulants are calculated using the analytical formula for Skellam distribution. \\ \\
The central values of $C_{6}$ for sample size $10^{5}$ events and $10^{7}$ events are found to be $-116.656$ and $-2.201$, which is interesting due to their $-ve$ signs. Though the theoretical prediction for a crossover/chiral phase transition suggests a $-ve$ value of $C_{6}$ to be a possible hint for such a transition~\cite{negc6:friman}, we find that the $-ve$ sign of $C_{6}$ (also $-ve$ sign of $C_{5}$ and $C_{7}$) in our simulation results is due to insufficient statistics of the sample set used. In some cases large values of $C_{4}$ are also observed. In the simulation, Skellam distribution of net-proton is generated using mean ($C_{1}$) of experimentally obtained proton and antiproton distributions. Hence, we would like to emphasize that in order to extract information related to unique physics features through the higher order cumulant studies, the first step is to ensure that the sample has sufficient statistics. 
\begin{figure}[hbtp]
\includegraphics[scale=0.45]{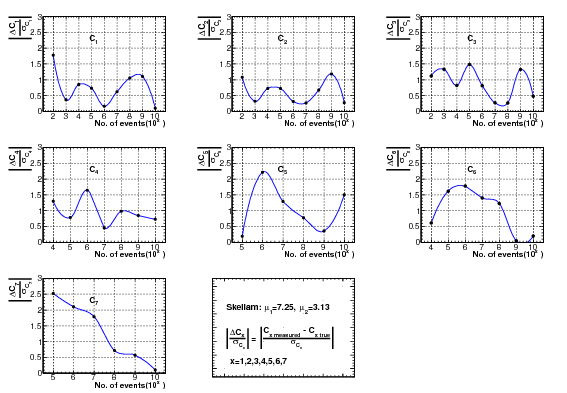}
\caption{(Color online) Sample size dependence of standard deviation (Modulus deviation of cumulants from their true values scaled with $1\sigma$ statistical error) of cumulants up to $7^{th}$ order in the Poissonian model with input parameters from proton and antiproton distribution in most central (0-5\%) Au$+$Au collision at \sNN$=62.4$~GeV.}
\label{fig:skell_chisq_62G} 
\end{figure}

We have performed similar study by taking inputs from another RHIC collision energy $i.e.$ for \sNN$=62.4$~GeV~\cite{starNP} (where the difference in mean values of proton and antiproton are larger than that for $200$~GeV) as shown in Fig.~\ref{fig:band_sklm_62G}. As seen in the case of \sNN$=200$~GeV, cumulants calculated in simulation randomly fluctuate around the true value and with increase in sample size better agreement between cumulants with their true values is observed. While the central values of $C_{6}$ and $C_{7}$ pick up $- ve$ values for sample sizes of both $10^{5}$ and $10^{7}$ events, the $\pm1\sigma$ statistical error makes the values consistent with true values. Figure ~\ref{fig:skell_chisq_62G} shows that cumulants lie in a range of $0.04\sigma$ to $2.5\sigma$.
\begin{figure*}[hbtp]
\centering 
\includegraphics[scale=0.3]{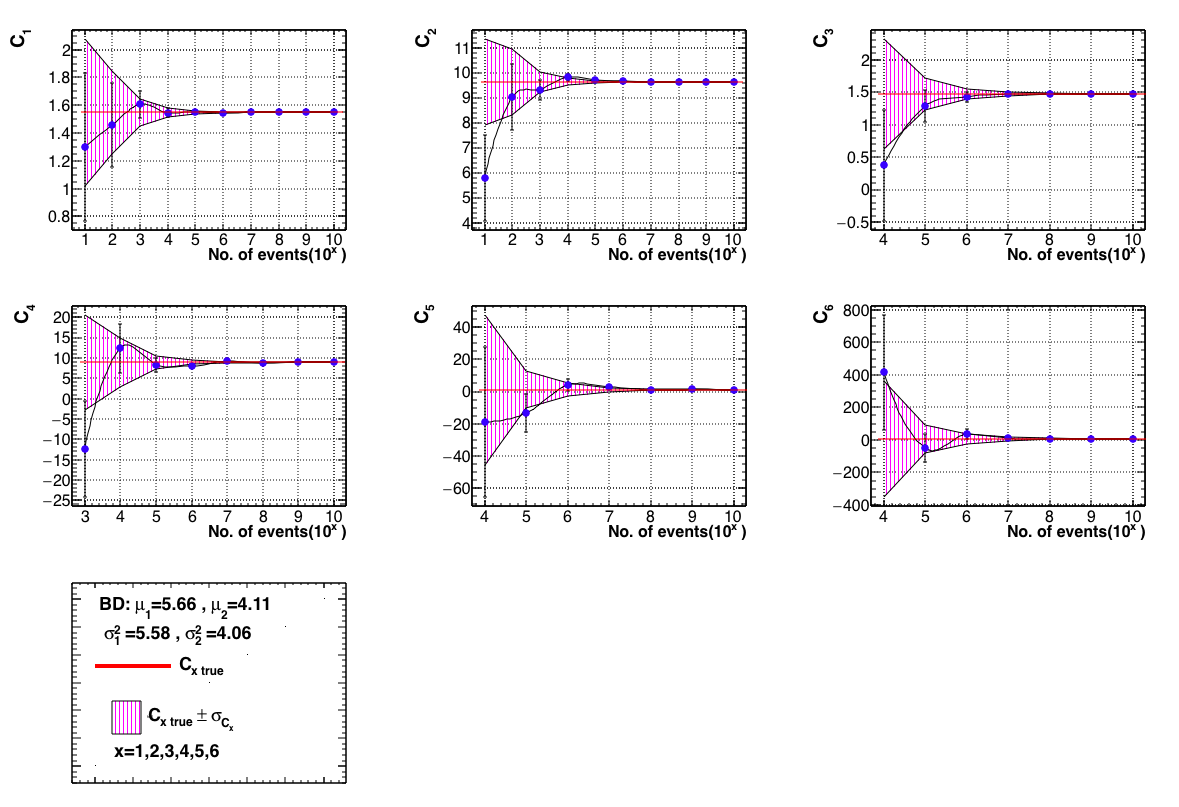}
\caption{(Color online) Cumulants of net-proton distribution upto $6^{th}$ order are shown as function of sample size. The net-proton distribution is difference of binomially distributed proton and antiproton with input parameters $\mu_{1}$, $\sigma_{1}^{2}$ (for proton) and $\mu_{2}$, $\sigma_{2}^{2}$ (for antiproton) same as corresponding experimental value in most central (0-5\%) Au$+$Au collision at \sNN$=200$~GeV. Blue solid circle markers represent the calculated cumulant values from simulated net-proton distribution. The solid red line represents the true value of cumulants obtained using analytical formula for a binomial distribution. The magenta band represents $\pm1\sigma$ statistical error (from delta theorem method) about the true value of cumulants.}
\label{fig:band_bino_200G} 
\end{figure*}
\begin{figure*}[hbtp]
\centering 
\includegraphics[scale=0.3]{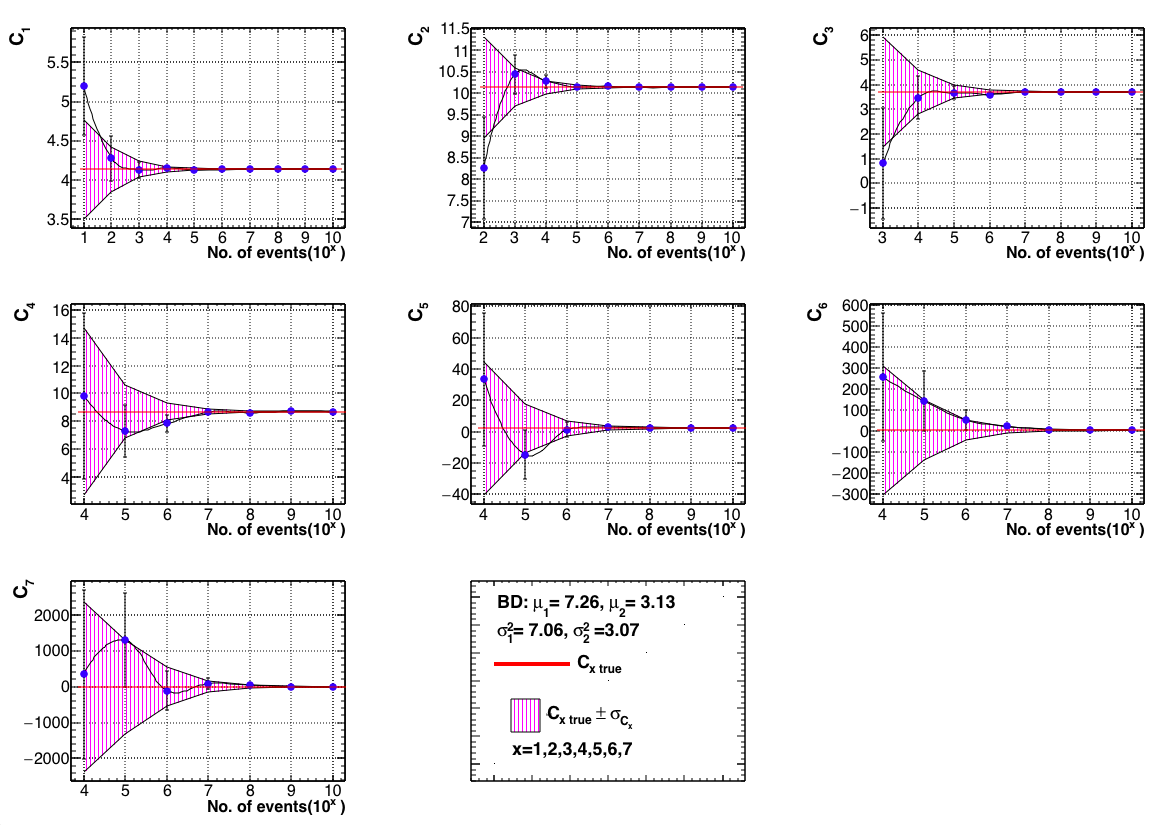}
\caption{(Color online) Cumulants of net-proton distribution upto $7^{th}$ order are shown as function of sample size. The net-proton distribution is difference of binomially distributed proton and antiproton with input parameters $\mu_{1}$, $\sigma_{1}^{2}$ (for proton) and $\mu_{2}$, $\sigma_{2}^{2}$ (for antiproton) same as corresponding experimental value in most central (0-5\%) Au$+$Au collision at \sNN$=62.4$~GeV. Blue solid circle markers represent the calculated cumulant values from simulated net-proton distribution. The solid red line represents the true value of cumulants obtained using analytical formula for a binomial distribution. The magenta band represents $\pm1\sigma$ statistical error (from delta theorem method) about the true value of cumulants.}
\label{fig:band_bino_62G} 
\end{figure*}

In the binomial model, net-proton distribution is generated from difference of binomially distributed proton and antiproton where input parameters $\mu_{1}$, $\sigma_{1}^{2}$ is the efficiency corrected mean and variance of proton distribution and $\mu_{2}$, $\sigma_{2}^{2}$ is the efficiency corrected mean and variance of antiproton distribution obtained in the experiment. The results from Binomial model are shown in Fig.~\ref{fig:band_bino_200G} and ~\ref{fig:band_bino_62G}. For \sNN$=200$~GeV we show results for only up to $6^{th}$ order cumulants\footnote{True value of $7^{th}$ order cumulant for \sNN$=200$~GeV in Binomial model is 0.14. The relative error ($|\frac{C_{meas.}-C_{true}}{C_{true}}|$) takes up large value because of vanishingly small value of the denominator.}. For both \sNN$=200$~GeV and $62.4$~GeV, qualitative behaviour of cumulants as a function of sample size is similar as seen in Poissonian model.
\begin{figure}[hbtp]
\includegraphics[scale=0.45]{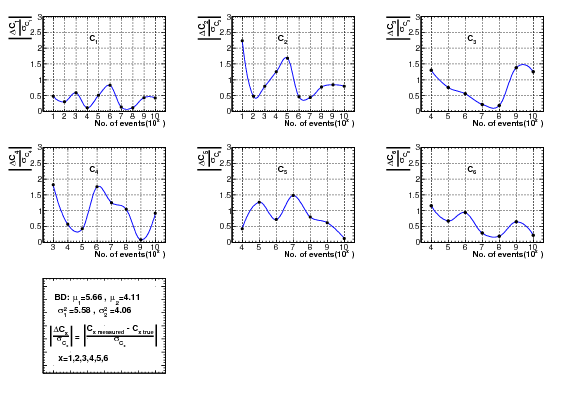}
\caption{(Color online) Sample size dependence of standard deviation (Modulus deviation of cumulants from their true values scaled with $1\sigma$ statistical error) of cumulants up to $6^{th}$ order in the binomial model with input parameters from proton and antiproton distribution in most central (0-5\%) Au$+$Au collision at \sNN$=200$~GeV.}
\label{fig:bino_chisq_200G} 
\end{figure}
In the binomial model, for \sNN$=200$~GeV, $-ve$ values are found for $C_{4}$ (for sample size $10^{3}$), $C_{5}$ (for sample size $10^{4}$ and $10^{5}$) and $C_{6}$ (for sample size $10^{5}$). For \sNN$=62.4$~GeV, $C_{6}$ shows $+ve$ values across different sample sizes in this model. 
\begin{figure}[hbtp]
\includegraphics[scale=0.45]{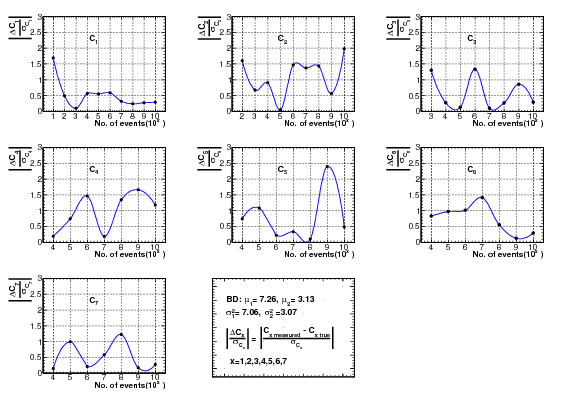}
\caption{(Color online) Sample size dependence of standard deviation (Modulus deviation of cumulants from their true values scaled with $1\sigma$ statistical error) of cumulants up to $7^{th}$ order in the binomial model with input parameters from proton and antiproton distribution in most central (0-5\%) Au$+$Au collision at \sNN$=62.4$~GeV.}
\label{fig:bino_chisq_62G} 
\end{figure}
Figure~\ref{fig:bino_chisq_200G} and ~\ref{fig:bino_chisq_62G} show cumulants agree with their true values with absolute standard deviation varying in a range of $0.07\sigma$ to $2.2\sigma$ for  \sNN$=200$~GeV and $0.04\sigma$ to $2.3\sigma$ for \sNN$=62.4$~GeV, respectively. As pointed out in the text before, settling down on proper statistically sound sample size is very important before interpretation of magnitude or sign of experimentally measured the higher order cumulants.\\
As can be seen from the above results and discussions and the fact that any experiment in laboratory will always have a finite sample size, proper estimation of statistical error is very crucial. In next part of current section, we will investigate the results from the study of different methods used to estimate the statistical error on various order cumulants.
\subsubsection{\label{sec:ErrorResults}Comparison of methods of error estimation}
\begin{figure}[ht!]
\centering 
\includegraphics[scale=0.45]{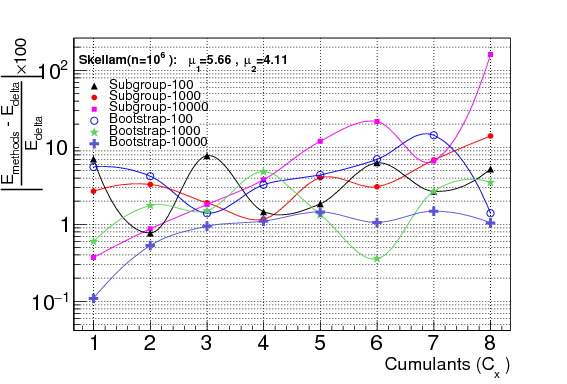}
\caption{(Color online) Relative percentage deviation of errors obtained in different methods (Sub-group and Bootstrap) from those obtained in delta theorem method is shown for different cumulants upto $8^{th}$ order for a Skellam distribution of sample size $10^{6}$ generated using input parameters $\mu_{1}$ (and $\mu_{2}$) taken from experimental value of means ($C_{1}$) of proton (and antiproton) distributions in most central (0-5\%) Au$+$Au collision at \sNN$=200$~GeV. Black triangle, red solid circle and magenta solid square markers correspond to cases of $(i)\,100$, $(ii)\,1000$ and $(iii)\,10000$ number of sub-samples respectively in the sub-group method. Open blue circle, green star and purple cross symbols represent bootstrap method with number of bootstrap samples corresponding to cases $(i)$,$(ii)$ and $(iii)$, respectively.}
\label{fig:compdelta_skellam200G_size6}
\end{figure}

Figure~\ref{fig:compdelta_skellam200G_size6} shows the relative deviation of errors in percentage obtained from bootstrap and sub-group methods from those obtained in delta theorem for Skellam distributed net-proton with a sample size of $10^{6}$ events. The input parameters for Skellam distribution are taken from experimental data for most central (0-5\%) Au$+$Au collision at \sNN$=200$~GeV. In bootstrap method errors are calculated for 3 different cases with $(i)\,100$, $(ii)\,1000$ and $(iii)\,10000$ numbers of bootstrap samples. Similarly, in the sub-group method errors are obtained with $(i)\,100$, $(ii)\,1000$ and $(iii)\,10000$ numbers of sub-samples. We observe in Fig.~\ref{fig:compdelta_skellam200G_size6} that the errors from sub-group and bootstrap methods are in good agreement with those from delta theorem for lower order cumulants. With increase in the order of cumulants, the relative percentage deviation also increases. It is important to note that the sub-group method with sub-samples $s=10000$ shows a maximum deviation of nearly $150$\% w.r.t. to delta theorem method. Errors obtained from the bootstrap method for $B=10000$ show close agreement with the error obtained from the delta theorem. 
\begin{figure*}[ht!]
\includegraphics[scale=0.27]{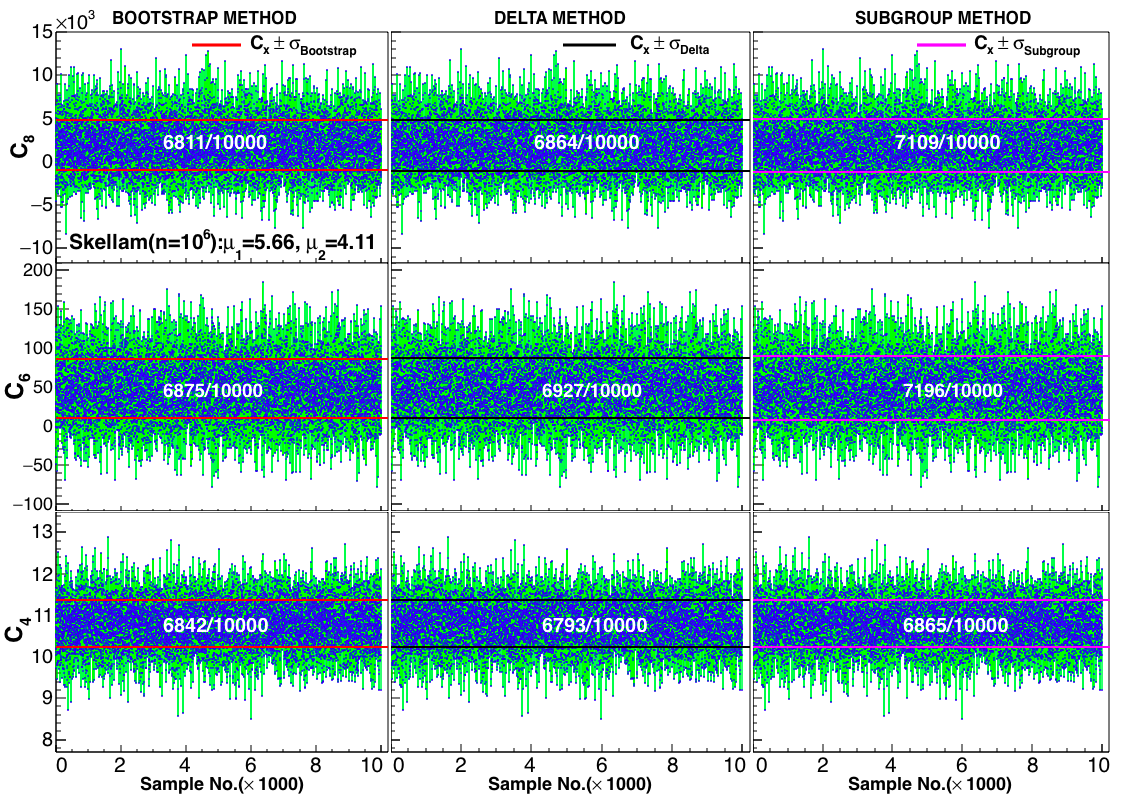}
\caption{(Color online) Verification of Gaussian (probability distribution) behaviour of errors obtained in different methods in a sample of size $10^{6}$. Values of cumulants $C_{4}$, $C_{6}$ and $C_{8}$, shown in blue points are calculated for 10000 samples generated by employing bootstrap sampling method from a Skellam distributed parent sample (of size $10^{6}$) characterized by $\mu_{1}$ (and $\mu_{2}$) taken from experimental value of means ($C_{1}$) of proton (and antiproton) distributions in most central (0-5\%) Au$+$Au collision at \sNN$=200$~GeV. Red lines in left most column panel figures represent bootstrap error on cumulants in parent sample. Black lines in middle column figures represent delta method error on cumulants. Magenta lines in right most column figures represent error on cumulants from sub-group method.}
\label{fig:ver_err_skellam200G_size6}
\end{figure*}
\begin{figure*}[ht!]
\centering 
\includegraphics[scale=0.28]{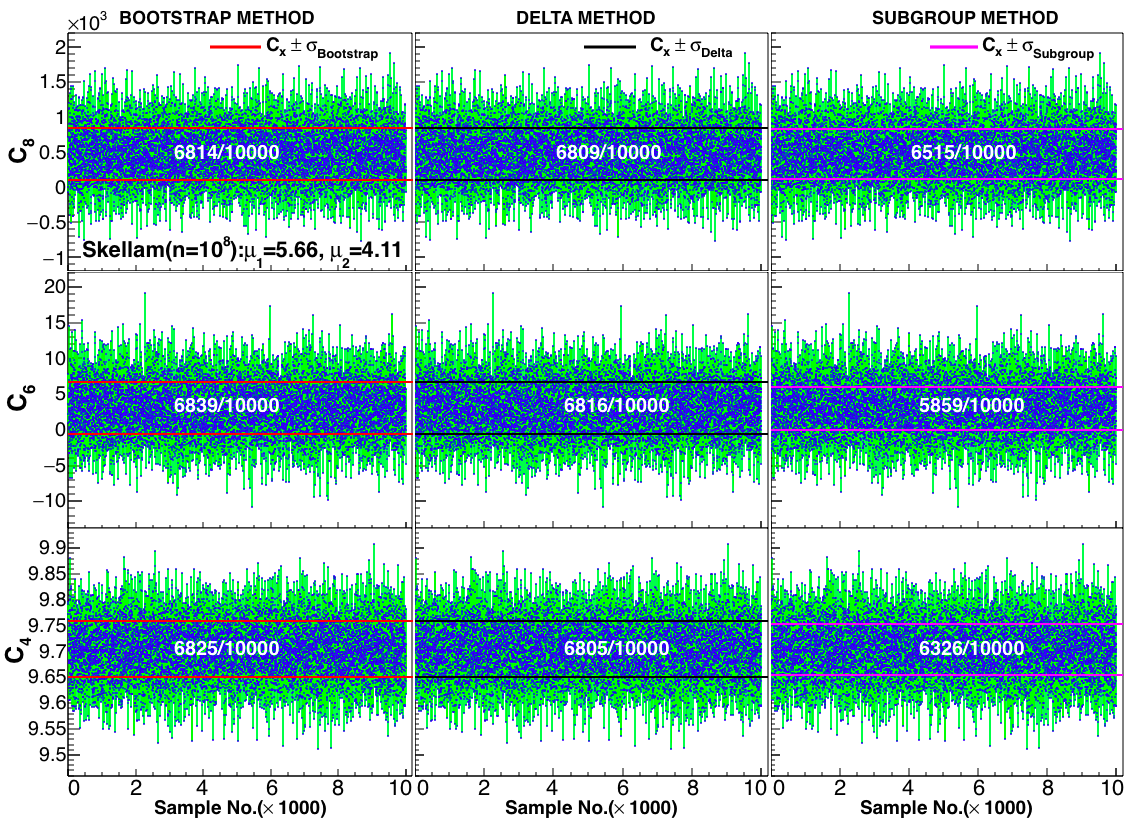}
\caption{(Color online) Verification of Gaussian (probability distribution) behaviour of errors obtained in different methods in a sample of size $10^{8}$. Values of cumulants $C_{4}$, $C_{6}$ and $C_{8}$, shown in blue points are calculated for 10000 samples generated by employing bootstrap sampling method from a Skellam distributed parent sample (of size $10^{8}$) characterized by $\mu_{1}$ (and $\mu_{2}$) taken from experimental value of means ($C_{1}$) of proton (and antiproton) distributions in most central (0-5\%) Au$+$Au collision at \sNN$=200$~GeV. Red lines in left most column panel figures represent bootstrap error on cumulants in parent sample. Black lines in middle column figures represent delta method error on cumulants. Magenta lines in right most column figures represent error on cumulants from sub-group method.}
\label{fig:ver_err_skellam200G_size8} 
\end{figure*}
To test the Gaussian (probability distribution) behaviour of errors, in Fig.~\ref{fig:ver_err_skellam200G_size6}, we show the values $C_{4}$, $C_{6}$ and $C_{8}$ calculated for $10000$ samples constructed from the same parent sample as used in Fig.~\ref{fig:compdelta_skellam200G_size6}. The replicated samples (each of size $10^{6}$), used in this study are constructed by random selection of entries from parent sample with replacement. We find that within $\pm1\sigma$ of bootstrap error $68$\% of the sampling distribution lies for all the cumulants shown. Delta theorem method also satisfies the Gaussian behaviour as $67.9$\%, $69.2$\% and $68.6$\% of the total number of samples lie within $\pm1\sigma$ delta method error for $C_{4}$, $C_{6}$ and $C_{8}$ respectively. For $C_{6}$ and $C_{8}$ sub-group method is seen to over estimate the errors.
\begin{figure}[ht!]
\centering 
\includegraphics[scale=0.45]{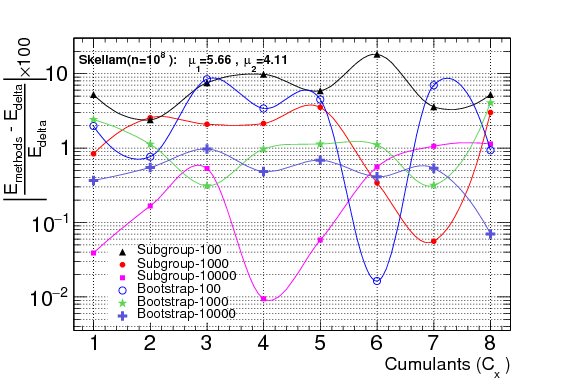}
\caption{(Color online) Relative percentage deviation of errors obtained in different methods (Sub-group and Bootstrap) from those obtained in delta theorem method is shown for different cumulants upto $8^{th}$ order for a Skellam distribution of sample size $10^{8}$ generated using input parameters $\mu_{1}$ (and $\mu_{2}$) taken from experimental value of means ($C_{1}$) of proton (and antiproton) distributions in most central (0-5\%) Au$+$Au collision at \sNN$=200$~GeV. Black triangle, red solid circle and magenta solid square markers correspond to cases of $(i)\,100$, $(ii)\,1000$ and $(iii)\,10000$ number of sub-samples respectively in the sub-group method. Open blue circle, green star and purple cross symbols represent bootstrap method with number of bootstrap samples corresponding to cases $(i)$,$(ii)$ and $(iii)$, respectively.}
\label{fig:compdelta_skellam200G_size8}
\end{figure}
Figure~\ref{fig:compdelta_skellam200G_size8} shows the relative percentage deviation of errors from sub-group and bootstrap methods from those obtained using delta theorem method for different order of cumulants (up to $8^{th}$ order) in Skellam distributed net-proton sample of size $10^{8}$ events. Only by changing the parent sample size to $10^{8}$, we observe an overall better agreement between errors from different methods compared to the case with a parent sample size of $10^{6}$ events. In the current case, the sub-group method with the number of sub-samples, $s=10000$ shows better agreement (agrees nearly within $1$\%) with the delta theorem. The bootstrap method with $B=10000$ also shows agreement consistently within $1$\% with delta theorem method across all order of cumulants shown. Values $C_{4}$, $C_{6}$ and $C_{8}$ calculated for $10000$ samples with $\pm1\sigma$ errors on them calculated from different methods are shown in Fig.~\ref{fig:ver_err_skellam200G_size8}. For all order cumulants shown, $\pm1\sigma$ error from bootstrap and delta theorem method include $68$\% of sampling distributions of cumulants. In this case, out of $10000$ samples $63$\% for $C_{4}$, $58$\% for $C_{6}$ and $65$\% for $C_{8}$ fall within $\pm1\sigma$ error from sub-group method. From above studies, we observe that sub-group method does not satisfy the Gaussian behaviour of cumulants across sampling distributions.

\section{\label{sec:discussions}Discussions}
In the context of measurement of higher order cumulants, we discuss below few points on the results mentioned in the previous section.

Cumulant values in both Poissonian and binomial model with input parameters taken from measured event-by-event distributions of proton and antiprotons for Au+Au collisions at \sNN$=200$ and $62.4$~GeV show statistical randomness as a function of sample size, saturates and approaches their true values towards the large value of sample size. 

\begin{figure}[ht!]
\centering 
\includegraphics[scale=0.3]{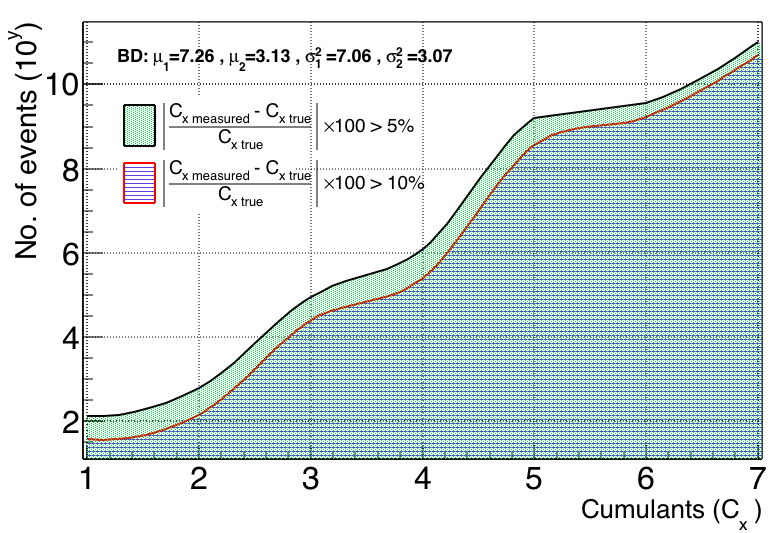}
\caption{(Color online) Exclusion limits on event statistics for measurement of cumulants of net-proton. The net-proton distribution is difference of binomially distributed proton and antiproton with input parameters $\mu_{1}$, $\sigma_{1}^{2}$ (for proton) and $\mu_{2}$, $\sigma_{2}^{2}$ (for antiproton) same as corresponding experimental values in most central (0-5\%) Au$+$Au collision at \sNN$=62.4$~GeV. Case - $(i)$: Red bottom line corresponds to minimum statistics needed for performing measurement of various order cumulants up to $7^{th}$ order in 10\% relative deviation from the true value. Case - $(ii)$: Black top line corresponds to the 5\% relative deviation from the true value. In the measurement of cumulants, filled regions below red and black lines are excluded for requiring better event statistics as described in cases $(i)$ and $(ii)$ respectively.}
\label{fig:bino62G_exclusion} 
\end{figure}

Higher order cumulants which characterize the subtle details of the shape of a statistical distribution are predicted to carry signals of phase transition of matter created in heavy-ion collisions. The $-ve$ sign of $C_{6}$ which is attributed to the physics of nature of phase transition in QCD phase diagram, is observed in our study for a statistically generated distribution where none of the above physics processes are included. Hence, this observation could only be attributed to the cumulants obtained from a sample with insufficient number of events. Hence obtaining a sample with sufficient statistics is very crucial for the experiments looking for signals of phase transition and critical point. \\
\begin{figure}[hbtp]
\centering 
\includegraphics[scale=0.3]{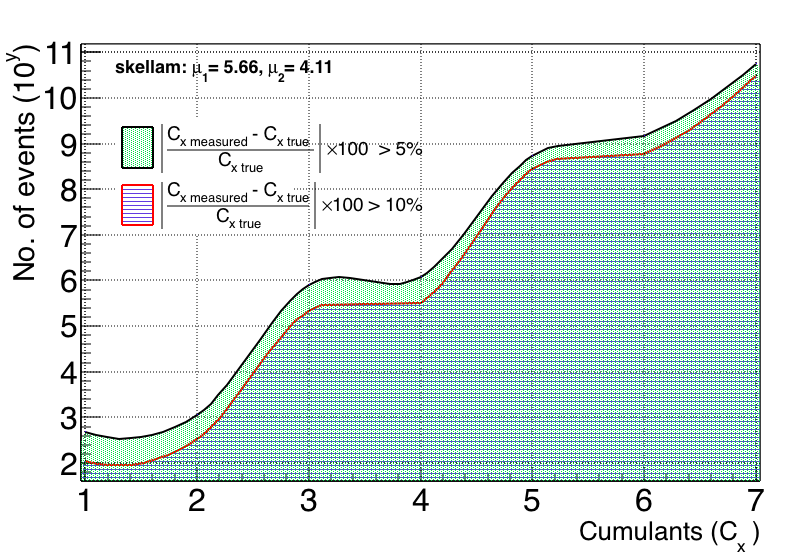}
\caption{(Color online) Exclusion limits on event statistics for measurement of cumulants of net-proton (Skellam distributed). Net-proton distribution is characterized by $\mu_{1}$ (and $\mu_{2}$) taken from experimental value of means ($C_{1}$) of proton (and antiproton) distributions in most central (0-5\%) Au$+$Au collision at \sNN$=200$~GeV. Case - $(i)$: Red bottom line corresponds to minimum statistics needed for performing measurement of various order cumulants up to $7^{th}$ order in 10\% relative deviation from the true value. Case - $(ii)$: Black top line corresponds to the 5\% relative deviation from the true value. In the measurement of cumulants, filled regions below red and black lines are excluded for requiring better event statistics as described in cases $(i)$ and $(ii)$ respectively.}
\label{fig:skellam200G_exclusion} 
\end{figure}
\begin{figure}[hbtp]
\centering 
\includegraphics[scale=0.3]{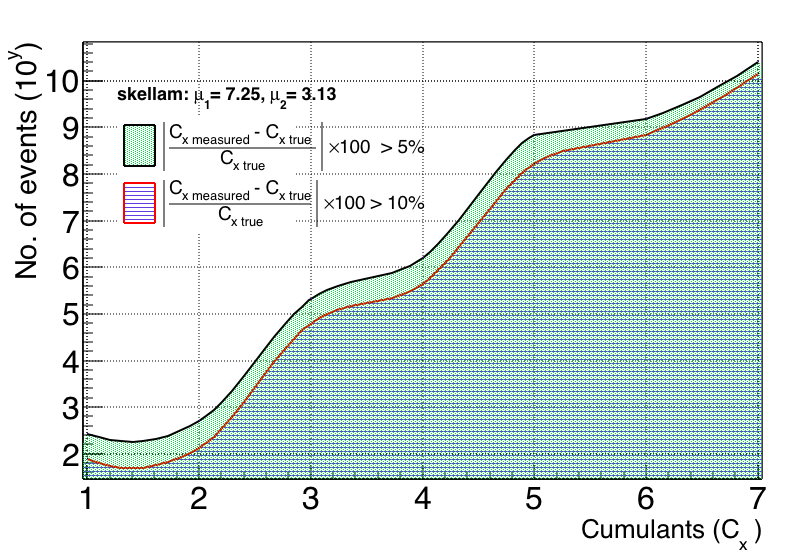}
\caption{(Color online) Exclusion limits on event statistics for measurement of cumulants of net-proton (Skellam distributed). Net-proton distribution is characterized by $\mu_{1}$ (and $\mu_{2}$) taken from experimental value of means ($C_{1}$) of proton (and antiproton) distributions in most central (0-5\%) Au$+$Au collision at \sNN$=62.4$~GeV. Case - $(i)$: Red bottom line corresponds to minimum statistics needed for performing measurement of various order cumulants up to $7^{th}$ order in 10\% relative deviation from the true value. Case - $(ii)$: Black top line corresponds to the 5\% relative deviation from the true value. In the measurement of cumulants, filled regions below red and black lines are excluded for requiring better event statistics as described in cases $(i)$ and $(ii)$ respectively.}
\label{fig:skellam62G_exclusion}
\end{figure}
In order to guide the experiments on what could be sufficient statistics using our model (Poissonian and Binomial) based study, we have obtained the exclusion limits on statistics of data sample for precise measurement of cumulants with two different input parameters. We assume that the signal (due to phase transition or QCD critical point) to be at the level of 5\% or 10\% above the statistical level for each order of cumulant (for simplicity). Figure~\ref{fig:skellam200G_exclusion} shows the exclusion curves on sample statistics for various order cumulants in the Poissonian model with input parameters fixed from measured proton and antiproton distributions in Au+Au collisions at \sNN$=200$~GeV. The curves (black and red) represent the minimum statistics needed for measurement of cumulants with certain required precision ($5$\% and $10$\% relative deviation from their true values). This also translates into the minimum statistics needed to observe a signal of strength of 5\% or 10\% above the statistical baseline. The statistics regions lying above the curves are desirable for measurement of cumulants for a particular order, while regions below the curve are excluded by the requirement of the above-mentioned precision. For example, estimation of $C_{4}$ within an accuracy of $5$\% and $10$\% require a minimum statistics of $1.2$ million and $0.3$ million events respectively. Similarly for $C_{6}$, accuracy of $5$\% and $10$\% need minimum statistics of $1.5$ billion and $588$ million events respectively. In Fig.~\ref{fig:skellam62G_exclusion}, we show similar exclusion curve for Poissonian model with input parameters of \sNN$=62.4$~GeV. For measurement of $C_{4}$, $1.6$ million (for $5$\%) and $0.4$ million (for $10$\%) minimum event statistics is needed. For $C_{6}$, accuracy of $5$\% and $10$\% require a minimum statistics of $1.5$ billion and $684$ million events.
\begin{figure}[hbtp]
\centering 
\includegraphics[scale=0.3]{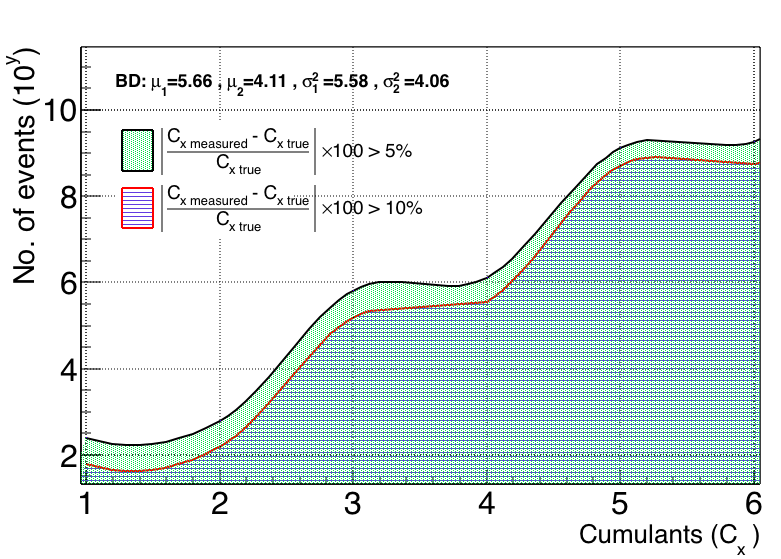}
\caption{(Color online) Exclusion limits on event statistics for measurement of cumulants of net-proton. The net-proton distribution is difference of binomially distributed proton and antiproton with input parameters $\mu_{1}$, $\sigma_{1}^{2}$ (for proton) and $\mu_{2}$, $\sigma_{2}^{2}$ (for antiproton) same as corresponding experimental values in most central (0-5\%) Au$+$Au collision at \sNN$=200$~GeV. Case - $(i)$: Red bottom line corresponds to minimum statistics needed for performing measurement of various order cumulants up to $6^{th}$ order in 10\% relative deviation from the true value. Case - $(ii)$: Black top line corresponds to the 5\% relative deviation from the true value. In the measurement of cumulants, filled regions below red and black lines are excluded for requiring better event statistics as described in cases $(i)$ and $(ii)$ respectively.}
\label{fig:bino200G_exclusion}
\end{figure}
Figure~\ref{fig:bino200G_exclusion} and Fig.~\ref{fig:bino62G_exclusion} show the exclusion curves for minimum statistics required for precision measurement of cumulants in a binomial model for input parameters for both \sNN$=200$ and $62.4$~GeV. In the case of \sNN$=200$~GeV for measurement of $C_{4}$ event sample with size of $1.3$ million (for $5$\% accuracy) and $0.3$ million (for $10$\% accuracy) and for measurement of $C_{6}$ event sample with size of $1.8$ billion (for $5$\% accuracy) and $552$ million (for $10$\% accuracy) are required. Similarly in the case of \sNN$=62.4$~GeV, measurement of $C_{4}$ requires event sample with size of $1.2$ million (for $5$\% accuracy) and $0.2$ million (for $10$\% accuracy) and measurement of $C_{6}$ requires sample size of $3.6$ billion (for $5$\% accuracy) and $1.6$ billion (for $10$\% accuracy). We observe that, for higher ($C_{5}$ and higher order) cumulants, the binomial model needs larger statistics compared to Poissonian model to measure cumulants within similar degrees of precision. This is important to notice that, the required event statistics reported above is for a single centrality bin (say $e.g.$ 0-5\%). So, in one collision energy for precise measurement of higher order cumulants such as $C_{4}$ and $C_{6}$, the total number of events (in full centrality classes) would be at least $1$ order of magnitude higher than those quoted above for a single centrality bin. This size of statistics, in fact, is very large compared to the currently available experimental data statistics.
\\Though the statistics of data sample required are very large for measuring higher cumulants within $5$\% and $10$\% of their true values, within $\pm1\sigma$ statistical error most of the cumulants agree with their true values for the events statistics available for current experimental data. However, a certain degree of caution should be taken when interpreting the energy dependence, centrality dependence and sign of the higher order cumulants. Till the experiments run for longer period and accumulate enough statistics for these measurements, one should obtain the statistical errors carefully and look for systematic variations across centrality/energy for physical interpretation.
\section{\label{sec:summary}Summary}
Higher order cumulants of distributions of conserved quantities are predicted to carry signals of phase transitions. The higher order cumulants are sensitive to the shape of the distributions and their values are driven by tails of the distributions. Hence the biggest challenges in such measurements are two fold (a) limited statistics and (b) correct estimation of statistical errors. We tried to address both these issues in this work. We study via a Monte Carlo simulation method, the effect of limited data statistics on the accuracy of the estimation of different order cumulants in two different models and for two different experimental input conditions. We observe that the estimated values of the higher order cumulants have a strong dependence on sample size and central values for cumulants randomly fluctuate around their true values at low event statistics, saturate and approach their true values for the large size of event sample. $C_{6}$ of net-proton distribution is predicted to show negative values if the system formed in heavy-ion collisions freezes-out close to a crossover transition. In this simulation, which has no phase transition effects incorporated, $C_{6}$ of the net-proton distribution (characterized only by lowest order cumulants of proton and antiproton distribution from experiment) shows the $-ve$ sign for some values of sample statistics. The positive point however is that, irrespective of the sample size the estimated values lies within $\pm1\sigma$ statistical error (obtained using Delta theorem method) on the cumulants from their true values, in most of the cases (for rest few cases agreements are observed within $\pm2.5\sigma$). In this work, we attribute this behaviour ($-ve$ sign of $C_{6}$) to purely statistical effect because simulated net-proton distribution does not include any physics of phase transitions. We have discussed the methods for accurate estimation of statistical error. We have studied three different methods namely, delta theorem method, the bootstrap method and sub-group method. We have verified the Gaussian probability distribution nature of the errors. We find bootstrap method is a robust method to estimate statistical error followed by the method using the delta theorem. Finally we have presented the exclusion limits on the minimum statistics of data sample needed for estimation of cumulants of different order in order to detect a signal (related to phase transition or critical point) of $5$\% and $10$\% above the statistical baseline using different models and input parameters.

\section*{\textbf{ACKNOWLEDGEMENTS}}
We thank Sourendu Gupta for useful discussions on error estimation methods for statistical distributions. We also would like to thank Nu Xu and Zhangbu Xu for helpful discussions on experimental aspects related to the measurement of higher cumulants with limited statistics. B.M. acknowledges the financial support from J C Bose National Fellowship of DST, Government of India. D.M. and A.P. acknowledge the financial support from Department of Atomic Energy, Government of India.
\appendix
\section{\label{sec:appA}Relation between cumulants and central moments}
Cumulants ($C_{n}$) in terms of central moments ($\mu_{n}$) are given by:
\begin{itemize}
	\item $C_2=\mu_2$
	\item $C_3=\mu_3$
	\item $C_4=\mu_4-3\mu_2^2$
	\item $C_5=\mu_5-10\mu_3\mu_2$
	\item $C_6=\mu_6-15\mu_4\mu_2-10\mu_3^2+30\mu_2^3$
	\item $C_7=\mu_7-21\mu_5\mu_2-35\mu_4\mu_3+210\mu_3\mu_2^2$
	\item $C_8=\mu_8-28\mu_6\mu_2-56\mu_5\mu_3-35\mu_4^2+420\mu_4\mu_2^2+560\mu_2\mu_3^2-630\mu_2^4$
\end{itemize}


\section{\label{sec:appB}Cumulants of binomial and Skellam distribution}
\subsubsection{\label{sec:appB1}Binomial distribution}
Binomial distribution is characterized by two parameters i.e.
\\Total number of trials: $n$ and 
\\Probability of success: $p$.
Atleast two cumulants from experiment are needed to fix above two parameters.
\\First eight cumulants of binomial distribution in terms of these parameters are given as:\\
\begin{itemize}
	\item $C_1=np$
	\item $C_2=np(1-p)$
	\item $C_3=np(1-3p+2p^2)$
	\item $C_4=np(1-7p+12p^2-6p^3)$
	\item $C_5=np(1-15p+50p^2-60p^3+24p^4)$
	\item $C_6=np(1-31p+180p^2-390p^3+360p^4-120p^5)$
	\item $C_7=np(1-63p+602p^2-2100p^3+3360p^4-2520p^5+720p^6)$
	\item $C_8=np(1-127p+1932p^2-10206p^3+25200p^4-31920p^5+20160p^6-5040p^7)$
	
\end{itemize}
\subsubsection{\label{sec:appB2}Skellam distribution}
Skellam distribution has two parameters, namely the means of two poisson variates, $\mu_1$ and $\mu_2$. The cumulants of a Skellam in terms of these two parameters are expressed as follows.\\
\begin{itemize}
	\item  $C_{2n+1}=\mu_1 - \mu_2$  for $n=0,1,2,3...$
	\item  $C_{2n}=\mu_1 + \mu_2$  for $n=1,2,3...$
\end{itemize}
All the even cumulants values are same and so are the odd ones.

\section{\label{sec:appC}Error on cumulants using delta theorem method}
Using delta theorem, errors obtained on various order cumulants are expressed as follows. To avoid the use of square root symbol repetitively, we write expressions for variances. 
\begin{itemize}
\item $\mathrm{Var}(C_1)=\mu_2/n$
\item $\mathrm{Var}(C_2)=(\mu_4-\mu_2^2)/n$
\item $\mathrm{Var}(C_3)=(\mu_6-\mu_3^2+9\mu_2^3-6\mu_2\mu_4)/n$
\item $\mathrm{Var}(C_4)=(\mu_8-12\mu_6\mu_2-8\mu_5\mu_3-\mu_4^2+48\mu_4\mu_2^2+64\mu_3^2\mu_2-36\mu_2^4)/n$
\item $\mathrm{Var}(C_5)=(\mu_{10}-\mu_5^2-10\mu_4\mu_6+900\mu_2^5-20\mu_3\mu_7-20\mu_8\mu_2+125\mu_2\mu_4^2+200\mu_4\mu_3^2-1000\mu_3^2\mu_2^2+160\mu_6\mu_2^2-900\mu_4\mu_2^3+240\mu_2\mu_3\mu_5)/n$
\item
$\mathrm{Var}(C_6)=(-30\mu_4\mu_8+510\mu_4\mu_2\mu_6+1020\mu_4\mu_3\mu_5+405\mu_8\mu_2^2-2880\mu_6\mu_2^3-9720\mu_3\mu_5\mu_2^2-30\mu_2\mu_{10}+840\mu_2\mu_3\mu_7+216\mu_2\mu_5^2-40\mu_3\mu_9+440\mu_6\mu_3^2-3600\mu_2^2\mu_4^2-9600\mu_2\mu_4\mu_3^2+13500\mu_4\mu_2^4+39600\mu_2^3\mu_3^2+\mu_{12}-\mu_6^2-12\mu_5\mu_7+225\mu_4^3-8100\mu_2^6-400\mu_3^4)/n$
\item $\mathrm{Var}(C_7)=(2590\mu_3\mu_4\mu_7+1890\mu_2\mu_4\mu_8-70\mu_3\mu_{11}-42\mu_2\mu_{12}+\mu_{14}-\mu_7^2+343\mu_2\mu_6^2-14\mu_6\mu_8+1911\mu_4\mu_5^2+558600\mu_2^2\mu_3^2\mu_4-76440\mu_2\mu_3\mu_4\mu_5-10584\mu_2^2\mu_5^2+299880\mu_2^3\mu_3\mu_5-1102500\mu_2^4\mu_3^2+861\mu_2^2\mu_{10}+176400\mu_2^3\mu_4^2-29400\mu_2^2\mu_4\mu_6+1715\mu_4^2\mu_6-14700\mu_3^2\mu_4^2-14700\mu_2\mu_4^3+79380\mu_2^4\mu_6+396900\mu_2^7-529200\mu_2^5\mu_4+1505\mu_3^2\mu_8+137200\mu_2\mu_3^4-15680\mu_3^3\mu_5+2310\mu_2\mu_3\mu_9-33600\mu_2^2\mu_3\mu_7-43120\mu_2\mu_3^2\mu_6-42\mu_5\mu_9+966\mu_2\mu_5\mu_7+2254\mu_3\mu_5\mu_6-10080\mu_2^3\mu_8-70\mu_4\mu_{10})/n$
\item $\mathrm{Var}(C_8)=(-56\mu_6\mu_{10}+4256\mu_3^2\mu_{10}+\mu_{16}-6350400\mu_2^8-4900\mu_4^4-112\mu_3\mu_{13}+1624\mu_2^2\mu_{12}+5040\mu_4^2\mu_8-71680\mu_3^3\mu_7+6272\mu_5^2\mu_6+512\mu_2\mu_7^2-26656\mu_2^2\mu_6^2-2399040\mu_2^5\mu_6+4480\mu_3\mu_6\mu_7-8467200\mu_2^2\mu_3^4+12700800\mu_2^6\mu_4+4704\mu_4\mu_6^2+940800\mu_3^4\mu_4-6174000\mu_2^4\mu_4^2+882000\mu_2^2\mu_4^3+1680\mu_2\mu_6\mu_8+322560\mu_2^4\mu_8-108360\mu_2^2\mu_4\mu_8-56\mu_2\mu_{14}+2007040\mu_2\mu_3^3\mu_5+9856\mu_4\mu_5\mu_7+59270400\mu_2^5\mu_3^2+6496\mu_3\mu_5\mu_8-140\mu_4\mu_{12}-75264\mu_3^2\mu_5^2-160160\mu_2\mu_3^2\mu_8-112\mu_5\mu_{11}+3808\mu_2\mu_5\mu_9-77952\mu_2^2\mu_5\mu_7-119840\mu_2^2\mu_3\mu_9-35280000\mu_2^3\mu_3^2\mu_4+2759680\mu_2^2\mu_3^2\mu_6-16\mu_7\mu_9+8960\mu_3\mu_4\mu_9-156800\mu_3\mu_4^2\mu_5+3684800\mu_2\mu_3^2\mu_4^2+-203840\mu_3^2\mu_4\mu_6+1340640\mu_2^3\mu_4\mu_6-15523200\mu_2^4\mu_3\mu_5+1626240\mu_2^3\mu_3\mu_7+677376\mu_2^3\mu_5^2+5600\mu_2\mu_4\mu_{10}-\mu_8^2-172480\mu_2\mu_4^2\mu_6-178752\mu_2\mu_4\mu_5^2-28560\mu_2^3\mu_{10}+5376\mu_2\mu_3\mu_{11}-257152\mu_2\mu_3\mu_5\mu_6+5597760\mu_2^2\mu_3\mu_4\mu_5-322560\mu_2\mu_3\mu_4\mu_7)/n$
\end{itemize}
where $n$ is the sample size.

\end{document}